\documentclass[11pt,a4paper]{article}
\usepackage{jheppub}
\pdfoutput=1

\usepackage[utf8]{inputenc}
\usepackage{amsmath}
\usepackage{amsfonts}
\usepackage{xcolor}
\usepackage{tikz}
\usetikzlibrary{arrows}
\usetikzlibrary{shapes}
\usepackage[english]{babel}
\usepackage[autostyle]{csquotes}
\usepackage{tabularx}

\newtheorem{conjecture}{Conjecture}

\newcommand{\CC}{\mathcal C}
\newcommand{\CH}{\mathcal H}

\newcommand{\CN}{\mathcal N}
\newcommand{\co}{\mathbb C}
\newcommand{\qu}{\mathbb H}
\newcommand{\q}{\mathsf{Q}}
\newcommand{\dif}{\mathsf{D}}
\newcommand{\h}{\mathsf{H}}
\newcommand{\iq}{\mathsf{I}}
\newcommand{\jq}{\mathsf{J}}
\newcommand{\hto}{\text{ht}(\mathcal O)}
\newcommand{\define}{\mathrel{\mathop:}=}
\newcommand{\be}{\begin{equation}} \newcommand{\ee}{\end{equation}}
\def\c3d#1{\mathcal{C}^{3d}(#1)}
\DeclareMathOperator{\rank}{\text{rank}}
\DeclareMathOperator{\tr}{\text{tr}}
\def\node#1#2{\overset{#1}{\underset{#2}{\circ}}}
\def\sqnode#1#2{\overset{#1}{\underset{#2}{\scriptstyle\square}}}
\def\flav#1{\overset{\scriptstyle#1}{\overset{\square}{\scriptstyle\vert}}}
\def\squ#1{\overset{{\scriptstyle\square}\rlap{\,\,$\scriptstyle#1$}}{\scriptstyle\vert}}
\def\squl#1{\overset{\llap{$\scriptstyle#1$\,\,}{\scriptstyle\square}}{\scriptstyle\vert}}
\def\upnode#1{\overset{\scriptstyle#1}{\overset{\displaystyle\circ}{\scriptstyle\vert}}}
\def\topnode#1#2{\overset{#1}{\overset{{\displaystyle\circ}\rlap{\,\,$\scriptstyle#2$}}{\scriptstyle\vert}}}

\def\CS{\mathcal S}
\def\lb{\left(}
\def\rb{\right)}
\def\sl#1{\mathfrak{sl}(#1,\mathbb{C})}
\def\so#1{\mathfrak{so}(#1,\mathbb{C})}
\def\sp#1{\mathfrak{sp}(#1,\mathbb{C})}
\def\bOr#1{\bar{\mathcal{O}}_{(#1)}}
\def\Or#1{{\mathcal{O}}_{(#1)}}
\def\Slo#1{\mathcal{S}_{(#1)}}
\def\Cb#1{\mathcal{C}\left(#1\right)}

\def\orbit#1#2{\overline{{\rm #1}_{#2}}}
\def\ba{\begin{equation} \begin{aligned}} \def\ea{\end{aligned}\end{equation}}

\preprint{Imperial/TP/18/AH/02}

\title{Quiver Subtractions}
\author[a]{Santiago Cabrera}
\author[b]{and Amihay Hanany}
\affiliation[a,b]{Theoretical Physics, The Blackett Laboratory\\
Imperial College London\\ SW7 2AZ United Kingdom}
\emailAdd{santiago.cabrera13@imperial.ac.uk}
\emailAdd{a.hanany@imperial.ac.uk}

 \abstract{We study the vacuum structure of gauge theories with eight supercharges. It has been recently discovered that in the Higgs branch of $5d$ and $6d$ SQCD theories with eight supercharges, the new massless states, arising when the gauge coupling is taken to infinity, can be described in terms of Coulomb branches of $3d \ \mathcal N=4$ quiver gauge theories. The description of this new phenomenon draws from the ideas discovered in the analysis of nilpotent orbits as Higgs and Coulomb branches of $3d$ theories and promotes the Higgs mechanism known as the \emph{Kraft-Procesi transition} to the status of a new operation between quivers. This is the \emph{quiver subtraction}. This paper establishes this operation formally and examines some immediate consequences. One is the extension of the physical realization of Kraft-Procesi transitions from the classical to the exceptional Lie algebras. Another result is the extension from \emph{special} nilpotent orbits to \emph{non-special} ones. One further consequence is the analysis of the effect in $5d\ \mathcal N=1$ SQCD of integrating out a massive quark while the gauge coupling remains infinite. In general, the subtraction of quivers sheds light on the different types of singularities within the Coulomb branch and the structure of the massless states that arise at those singular points; including the nature of the new Higgs branches that open up. This allows for a systematic analysis of mixed branches of $3d\ \mathcal N=4$ quivers that do not necessarily have a simple embedding in string theory. The subtraction of two quivers is an extremely simple resource for the theoretical physicist interested in the vacuum structure of gauge theories, and yet its power is so remarkable that is bound to play a crucial role in the coming discoveries of new and exciting physics in $5$ and $6$ dimensions.}


\begin{document}

\maketitle
\flushbottom

\section{Introduction}

The analysis of moduli spaces is a key step in the study of new physical phenomena arising in quantum field theories with eight supercharges\footnote{The myriad of examples goes from the study of monopole condensation and electric-magnetic duality  on $4d\ \mathcal N=2$ SYM \cite{SW94a,SW94b} to the very recent relation between the moduli space of $3d \ \mathcal N=4$ theories and the quantum Hall effect in $d=4+1$ dimensions \cite{BDLTT17}.}. In the past, the Higgs branch of theories with eight supercharges has not been studied with the same intensity as the Coulomb branch counterpart. This is due to the understanding that the classical Higgs branch does not receive quantum corrections \cite{APS96}. However in $5d$ and $6d$ theories this is not the full story. The Higgs branch of SQCD (with eight supercharges) at finite gauge coupling $g$ changes when the limit to infinite coupling is taken $1/g^2\rightarrow 0$.

In $5d$, the factor $1/g^2$ has scale of mass and new massless states arise at the limit. There is also a global symmetry enhancement \cite{Seiberg:1996bd, Morrison:1996xf, Intriligator:1997pq, Aharony:1997bh, DeWolfe:1999hj, Kim:2012gu, Bashkirov:2012re, Bao:2013pwa, Bergman:2013aca, Zafrir:2014ywa, Mitev:2014jza, Tachikawa:2015mha, Hayashi:2015fsa, Yonekura:2015ksa, Gaiotto:2015una, Bergman:2015dpa} and the appearance of new flat directions in the Higgs branch, that grows in dimension. In $6d$, the factor $1/g^2$ has the same scale as the tension of a string, and new tensionless strings appear at infinite coupling. In some cases the dimension of the Higgs branch grows by $29$ in a process called the \emph{small $E_8$ instanton transition}\footnote{See \cite{Mekareeya:2017sqh} for a recent application of this effect to the dynamics of M5-branes on ADE singularities.} \cite{Ganor:1996mu}.

The questions about the physical phenomena at infinite coupling could have been asked many years ago. The two main reasons that prevented this progress were the misconceptions about the unchanging nature of the Higgs branch and the fact that not enough tools were at the disposal of the theoretical physicist. However, with the advent of the non-perturbative studies of the Coulomb branch of $3d \ \mathcal N=4$ theories in terms of the \emph{monopole formula} \cite{CHZ13} and the introduction of nilpotent orbits in the description of Higgs and Coulomb branches of $3d \ \mathcal N=4$ theories\footnote{This is a generalization of the reduced moduli space of $1$ $A_n$ instanton on $\mathbb{C}^2$ (which is the Higgs branch of $3d\ \mathcal N = 4$ SQED with gauge group $U(1)$ and $n+1$ electrons) and the reduced moduli space of $1$ $D_n$ instanton on $\mathbb C^2$ 
(which is the Higgs branch of SQCD with gauge group $SU(2)$ and $n$ quarks) that constitute the starting point of the discussion in \cite{IS96}.} \cite{B70,KP79,Sl80,KP82,CM93,KS96,FH00,GW09,BTX10,CDT13,CHMZ14,Hanany:2016gbz,CH16,Cabrera:2017ucb,Hanany:2017ooe,CH17}, new discoveries have been possible in the last year. The first one is the description of the $5d$ Higgs branch at infinite coupling in terms of Coulomb branches of $3d$ quiver theories \cite{FHMZ17}. The second is the understanding of the small instanton transition in $6d$ in terms of \emph{Kraft-Procesi transitions}\footnote{See \cite{CH16} where these transitions are introduced as a specific type of Higgs mechanism in $3d\ \mathcal N=4$ quiver gauge theories with unitary nodes, and \cite{CH17} for an extension to orthogonal and symplectic gauge group factors. Note that in \cite{Heckman:2016ssk} Kraft-Procesi transitions between nilpotent orbits are used to characterize RG flows between $6d$ SCFTs, and the nilpotent orbit data is utilized in \cite{Mekareeya:2016yal} to study the anomalies and moduli spaces of the theories.}, also utilizing Coulomb branches of $3d \ \mathcal N=4$ theories \cite{HM18}.

A natural consequence of these two developments, and the advances that made them possible, is the promotion of the mathematical concept of \emph{transverse slice} \cite{B70,Sl80,KP81,KP82,FJLS15} to a new physical operation between quivers: the \emph{subtraction of quivers}. By subtracting $3d$ quivers the change in the $5d$ or $6d$ Higgs branch at infinite coupling can be understood in a remarkably simple way. This operation can be derived from the work in \cite{CH16} in unitary quivers and it was further extended in \cite{CH17} to \emph{orthosymplectic} quivers. In \cite{HM18} it was first employed to describe a $6d$ physical effect. The aim of this paper is to formally establish this operation and to examine some of its immediate consequences. In particular, we show how the study of mixed branches and different singularities within Coulomb branches of $3d\ \mathcal N=4$ quivers can greatly benefit from its application\footnote{The study of mixed branches has become relevant in the description of infra-red physics of \emph{bad} theories. The case of $3d\ \mathcal N=4$ SQCD with gauge group $U(N)$ and $N_f<2N-2$ quarks is studied in \cite{Assel:2017jgo}. \cite{Assel:2018exy} treats the SQCD theories with $Sp(N)$ gauge group. Quiver subtraction can simplify the computations of the mixed branches when the theories under study are more complicated $3d\ \mathcal N=4$ quivers.} (the information of mixed branches can be collected in this way into Hasse diagrams).

In Section \ref{sec:diff} we present a simple example in which the results of \cite{CH16} are recast in the language of quiver subtractions. Section \ref{sec:def} presents the formal definition of the \emph{quiver subtraction} operation. In Section \ref{sec:E6} we use the result to compute the Kraft-Procesi transitions between nilpotent orbits of exceptional algebras and also between \emph{non-special} nilpotent orbits of classical algebras, extending the physical realization presented in \cite{CH16,CH17}. In Section \ref{sec:5d} we use quiver subtractions to analyze the effect of integrating out a massive quark on the Higgs branch of $5d$ SQCD with eight supercharges. Section \ref{sec:con} collects some conclusions and future directions.

\section{The Subtraction of Two Quivers}\label{sec:diff}

This section illustrates with an example the idea of \emph{quiver subtraction}. This is a new way of interpreting the results that in \cite{CH16} were derived in the context of brane dynamics of Type IIB string theory. Let $\q$ be a $3d\ \mathcal N=4$ \emph{quiver}:
\be
	\q\define \ \node{}{1} - \node{\flav{1}}{2} - \node{\flav{1}}{2} - \node{}{1}
\ee
As is now conventional, this quiver represents a $3d \ \mathcal N=4$ theory with gauge group $G(\q)$ given by the circular nodes:
\be
	G(\q) = U(1)\times U(2)\times U(2)\times U(1)
\ee
The flavor group\footnote{Note that a global U(1) factor decouples, hence the symbol $S(...)$ on the definition of $F(\q)$.} $F(\q)$ is in turn determined by the square nodes:
\be
	F(\q) = S\lb U(1)\times U(1)\rb 
\ee
The edges of the quiver denote hypermultiplets of the theory transforming under the fundamental and antifundamental representations of the groups they connect\footnote{In particular, a hypermultiplet connecting two nodes with labels $a$ and $b$ divides into two half-hypers, one is in the irreducible representation with Dynkin labels $[1,0,\dots,0]_1\otimes [0,\dots,0,1]_{-1}$ of $U(a)\times U(b)$ and the other one in the complex conjugate $[0,\dots,0,1]_{-1}\otimes [1,0,\dots,0]_1$. }.
The moduli space of this theory has two special branches. The \emph{Coulomb branch} $\CC(\q)$ corresponds to a phase in which only scalar fields from vector multiplets acquire VEVs different from zero. In a generic non-singular point on the Coulomb branch the gauge group is broken to its maximal torus $U(1)^{6}$. This quiver has Coulomb branch:
\be 
	\CC(\q)=\bOr{2^2,1}\subset \mathfrak{sl}(5,\co)
\ee
where $\mathcal O_{(2^2,1)}$ is the \emph{next to minimal nilpotent orbit}\footnote{An introduction to \emph{nilpotent orbits} for physicists can be found in \cite[sec. 3.3]{CH16}, taken from the material in the book \cite{CM93}, which is the standard mathematical text on the topic. In this section we follow the same notation.} of $\mathfrak{sl}(5,\co)$. Its closure is called a \emph{nilpotent variety}, since it can be understood as the set of all $5\times 5$ matrices $M$ with complex entries that satisfy:
\begin{equation}
	\bOr{2^2,1} =\{M\in \co ^{5\times 5}|\tr (M)=0,\ M^2=0 \ \text{and}\  \rank (M)\leq 2  \}.
\end{equation}
The \emph{Higgs branch} $\CH(\q)$ corresponds to a phase in which only scalar fields from the hypermultiplets acquire non-zero VEVs. In a generic non-singular point of the Higgs branch the gauge group is fully broken. The Higgs branch of $\q$ is:
\be
	\CH(\q) = (\CS_{(3,2)}\cap \CN)\subset \mathfrak{sl}(5,\co)
\ee
Where $\Slo{3,2}$ is the \emph{Slodowy slice} transverse to the nilpotent orbit $\mathcal O_{(3,2)}\subset \sl 5$ and $\CN\subset \sl 5$ is the \emph{nilpotent cone}, that can be defined as the union of all nilpotent orbits:
\be
	\CN\define \bigcup_{\lambda\in \mathcal P(5)}\mathcal O _\lambda
\ee
where $\mathcal P(5)=\{(5),(4,1),(3,2),(2^2,1),(2,1^3),(1^5)\}$ is the set of all partitions of the integer $5$.
\paragraph{The Closure of a Nilpotent Orbit}
Let the rest of the section focus on the Coulomb branch $\CC(\q)$. The closure of a nilpotent orbit of $\sl n$ is the union:
\be 
	\bOr{\lambda} =\bigcup _{\lambda'\leq\lambda}\mathcal O_{\lambda ' }
\ee
where $\leq$ denotes the \emph{natural ordering} relation between partitions \cite{CM93}.
In this particular case, this means that:
\be
	\begin{aligned}
		\CC(\q) 	& = \bOr{2^2,1}\\
				& = \Or{2^2,1}\cup \Or{2,1^3}\cup \Or{1^5}\\
				& = \Or{2^2,1}\cup \bOr{2,1^3}
	\end{aligned}
\ee

The closure of the smaller orbit $\bOr{2,1^3}$ can also be realized as the Coulomb branch of a quiver gauge theory $\q'$:
\be 
	\CC(\q') = \bOr{2,1^3}\subset \sl 5
\ee
This quiver is:
\be 
	\q' \define \ \node{\flav 1}{1}-\node{}{1}-\node{}{1}-\node{\flav 1}{1}
\ee

\paragraph{The Subtraction of Two Quivers} Let us define the \emph{subtraction of quivers} as a new quiver $\dif = \q - \q'$:
\ba 
	\dif 	&= \q-\q'\\
		&= \ \lb\node{}{1}-\node{\flav 1}{2}-\node{\flav 1}{2}-\node{}{1} \rb - \lb\node{\flav 1}{1}-\node{}{1}-\node{}{1}-\node{\flav 1}{1} \rb \\
		&= \ \node{}{0}- \node{\flav 1}{1}-\node{\flav 1}{1}- \node{}{0} \\
		&= \ \node{\flav 1}{1}-\node{\flav 1}{1}
\ea
The ranks of the gauge nodes of $\dif$ are obtained by subtracting the ranks in $\q$ and $\q'$, while the flavor nodes are taken to be the same as the flavors in $\q$. The Coulomb branch of $\dif$ is also known:
\be 
	\CC(\dif) = a_2 
\ee
where $a_n$ is defined as the closure of the \emph{minimal nilpotent orbit}\footnote{The variety $a_n$ is also the reduced moduli space of $1$ $A_n$ instanton on $\mathbb{C}^2$. Notice that  the quiver $\dif$ is $3d$ mirror dual in the sense of \cite{IS96} to SQED with $3$ electrons. It is in this sense that we said before that the set of closures of nilpotent orbits $\bar{\mathcal O}_{\lambda}$, generalizes the moduli space of $A_n$ instantons when $\lambda$ is different from $(2,1^{n-1})$.} $\bOr{2,1^{n-1}}\subset \sl {n+1}$ (the notation is taken from \cite{KP82}). The relationship of $\CC (\dif)$ with $\Cb \q$ and $\Cb {\q'}$ is as follows: $a_2$ is the \emph{slice} in $\Cb \q= \bOr{2^2,1}$ \emph{transverse} to a point $x\in \Or{2,1^3}\subset \bOr{2^2,1}$ \cite{B70,Sl80,KP81,KP82}. This means that the singularity of $\Cb \q$ at $x \in {\Cb {\q'}}^\circ\subset \Cb \q$ is \emph{smoothly equivalent} to $\Cb \dif$ \cite[p.~9]{H14} (where $\Cb {\q'}^\circ$ denotes the \emph{interior} of the set of points in the Coulomb branch; it its important to restrict to the interior, since the transverse slice at other points could be larger, and the singularity would not be $\Cb D$).

\paragraph{New Massless States and Mixed Branches}

The physical implication is that locally in the Coulomb branch $\Cb \q$ two hypermultiplets become massless when the VEVs of the massless vector multiplets of $\q$ are restricted to be in the subspace ${\Cb{\q'}}^\circ\subset \Cb{\q}$. This opens up a new Higgs branch, which is equivalent to the Kleinian surface singularity $\CH(\dif ) = \mathbb{C}^2/\mathbb{Z}_3$. Hence, the appearance of new massless hypermultiplets and the transition to a mixed branch at a point $x\in {\Cb{\q'}}^\circ\subset\Cb \q$ is the same as that at the origin of the Coulomb branch of $\dif=\q-\q'$. The mixed branch is:
\be \label{eq:mix}
	\mathcal{B}(\q) = \Cb{\q'} \times \CH(\dif) 
\ee
The right hand side of equation (\ref{eq:mix}) should be understood as: \emph{the Coulomb part of the mixed branch $\mathcal B (\q)$ is isomorphic to $\Cb{\q'}$, while its Higgs branch is isomorphic to $\CH(\dif)$}. 
More generally, the appearance of singular transverse slices indicates that there are new massless states along the Coulomb branch at points that are not necessarily the origin of the branch. This generates the possibility for a phase transition to a mixed branch where the non-zero VEVs of matter hypermultiplets are fully defined by the Higgs branch of the \emph{subtraction quiver} $\dif$.

\section{General Definition}\label{sec:def}
Let two unitary quivers $\q$ and $\q'$ have the same number of gauge nodes and the same connections between said gauge nodes, such that $\Cb {\q'} \subseteq \Cb \q$ and:

\begin{enumerate}
	\item The ranks in the gauge nodes of $\q'$ are equal to the ranks of the gauge nodes of $\q$ except for a set of the gauge nodes in a connected subset of the quiver, in which the ranks of $\q'$ are strictly lower.
	\item The flavors attached to the subset of gauge nodes in $\q'$ with equal rank to $\q$ are the same to those attached in the corresponding nodes in $\q$ with one exception: there is an extra flavor attached to the gauge nodes adjacent to gauge nodes of $\q'$ with less rank than their corresponding nodes in $\q$.
	\item There is at least one flavor attached to one of the nodes in the subset of gauge nodes of $\q$ with higher rank than their counterpart in $\q'$.
\end{enumerate}

 Then, the \emph{subtraction quiver} $\dif = \q-\q'$ is defined as:
\begin{enumerate}
	\item $\dif$ is a quiver with the same number of gauge nodes as $\q$ or $\q'$ and the same connections between said gauge nodes.
	\item The rank of the gauge node in the $i$th position of $\dif$ is the difference between the rank of the $i$th gauge node of $\q$ and the rank of the $i$th gauge node of $\q'$.
	\item The flavor nodes of $\dif$ are in the same position and have the same labels as the flavor nodes on $\q$.
\end{enumerate}

\begin{conjecture}\label{con}
	If the Coulomb branch of $\q'$ is the closure $\Cb{\q'}=\bar{\mathcal O}$ where $\mathcal O$ is an orbit under the isometry group of $\Cb \q$, then the Coulomb branch $\Cb {\dif=\q-\q'}$ is the \emph{slice} in $\Cb \q$ \emph{transverse} to  $\mathcal O$. This means that there is a singularity on $\Cb \q$ at $x\in \mathcal O$. This singularity is smoothly equivalent to $\Cb \dif$. This implies that new hypermultiplets become massless at $x$ and a new Higgs branch opens up, which is isomorphic to $\CH (\dif)$, the Higgs branch of $\dif$. A transition is possible to a mixed branch of $\q$, $\mathcal B(\q) = \Cb{\q'}\times \CH (D)$ (by this we mean that the Coulomb part of the mixed branch is isomorphic to $\Cb{\q'}$ and its Higgs part is isomorphic to $\CH(\dif)$).
\end{conjecture}

This conjecture can be used to study the structure of the new massless states that appear along the Coulomb branch of any unitary $3d\ \mathcal N=4$ quiver $\q$ and the precise form of the new Higgs branches that open up.

 In particular, it can be used to analyse the Coulomb branches of $T_\rho(SU(n))$ theories \cite{GW09}, where $\Cb \q$ and $\Cb {\q'}$ are closures of nilpotent orbits of the same algebra $\sl n$ and their transverse slices are given by \cite{KP81,KP82} (this is a consequence of the analysis of the brane dynamics in \cite{CH16}). It can also be used to analyse vacuum of theories whose Coulomb branches are closures of nilpotent orbits of exceptional Lie algebras \cite{FJLS15}. These are new results, let us discuss them more explicitly in the next section.

\section{New Kraft-Procesi Transitions on Nilpotent Orbits}\label{sec:E6}

\subsection{Exceptional Algebras}

The definition of \emph{quiver subtraction} of Section \ref{sec:def} can be utilized to demonstrate the physical realization of Kraft-Procesi transitions between nilpotent orbits of exceptional algebras. Let us start with one example.

\paragraph{A Transverse Slice in $E_6$}

Let the closure of the \emph{next to minimal nilpotent orbit} of $\mathfrak{e}_6$ be denoted\footnote{In this section we deviate from the notation $\bar{\mathcal{O}}_\lambda$ for closures of nilpotent orbits, due to the fact that exceptional algebras do not have partitions $\lambda$ to label their different nilpotent orbits. Instead of the more conventional Bala-Carter labels \cite{CM93} used for exceptional algebras we have decided to adopt the notation ${\rm min}_G$, ${\rm n.min}_G$ and ${\rm n.n.min}_G$, for the \emph{minimal non trivial} nilpotent orbit of $Lie(G)$ and the two next orbits that are greater in dimension respectively. This decision is motivated by the fact that the discussion of this section is restricted to the smaller orbits of each exceptional algebra.} by $\orbit{n.min}{E_6}$. The $3d\ \mathcal N=4$ quiver $\q$ with Coulomb branch $\Cb \q=\orbit{n.min}{E_6}$ was found in \cite{Hanany:2017ooe}:
\be
	\q \define \ \node{\squ 1}{2} - \node{}{3} - \node{\topnode{}{2}}{4} - \node{}{3} - \node{\squ 1}{2}
\ee

On the other hand, the closure of the minimal nilpotent orbit of $\mathfrak{e}_6$, $\orbit{min}{E_6}$, is isomorphic to the reduced moduli space of $1$ $E_6$ instanton on $\mathbb{C}^2$. Therefore, the $3d\ \mathcal N=4$ quiver $\q'$ with Coulomb branch $\Cb {\q'}=\orbit{min}{E_6}$ is well known. $\q'$ is the extended Dynkin diagram of $E_6$, with ranks equal to the index of each node \cite{IS96}:

\be
	\q' \define \ \node{}{1} - \node{}{2} - \node{\topnode{\squ 1}{2}}{3} - \node{}{2} - \node{}{1}
\ee
The slice $S\subseteq \orbit{n.min}{E_6}$ transverse to ${\rm min}_{E_6}$ is computed to be a singular space, isomorphic to $a_5 =\orbit{min}{A_5}$ \cite{FJLS15}. According to Section \ref{sec:def}, the Coulomb branch of the subtraction quiver $\dif=\q-\q'$ should be the singular space $S=a_5$. Let us compute the subtraction:
\ba 
	\dif	&= \q-\q'\\
		&= \lb \node{\squ 1}{2} - \node{}{3} - \node{\topnode{}{2}}{4} - \node{}{3} - \node{\squ 1}{2}\ \ \rb - \lb \node{}{1} - \node{}{2} - \node{\topnode{\squ 1}{2}}{3} - \node{}{2} - \node{}{1} \rb \\
		&= \node{\squ 1}{1} - \node{}{1} - \node{\topnode{}{0}}{1} - \node{}{1} - \node{\squ 1}{1}\\
		&= \node{\squ 1}{1} - \node{}{1} - \node{}{1} - \node{}{1} - \node{\squ 1}{1}\\
\ea
Notice that $\q$ and $\q'$ fit into the definition of Section \ref{sec:def}, since the only gauge node of $\q'$ with the same rank as $\q$ (the top node with rank $2$) is adjacent to a node with smaller rank (the middle node with rank $3<4$), and it carries exactly one more flavor than its counterpart in $\q$. The resulting quiver $\dif$ is the $3d$ mirror quiver of SQED with $6$ electrons and eight supercharges. Its Coulomb branch is the reduced moduli space of $1$ $A_5$ instanton on $\mathbb{C}^2$ \cite{IS96} or, in the new language of nilpotent orbits, the closure of the minimal nilpotent orbit of $\sl 6$:
\be
	\Cb \dif = \CC \lb\node{\squ 1}{1} - \node{}{1} - \node{}{1} - \node{}{1} - \node{\squ 1}{1} \ \ \rb = a_5=\orbit{min}{A_5}
\ee

This implies that at a point $x\in {\rm min}_{E_6}\subset \Cb{\q}$ there are six hypermultiplets that become massless and a new Higgs branch opens up, equivalent to the Kleinian singularity $\CH (\dif)=\mathbb{C}^2/\mathbb{Z}_6$.

\begin{table}[t]
	\centering
	\begin{tabular}{|c|c|c|c|c|}
	\hline
	$i$ & $\Cb {\q_i}$ & $\q_i$ & $\dif_i$ & $\Cb {\dif_i}$ \\ \hline
	1 & $\overline{{\rm n.min}_{E_8}}$  & $\node{\squ 1}4-\node{}7-\node{\topnode{}5}{10}-\node{}8-\node{}6-\node{}4-\node{}2\ $ & $\node{\squ 1}2-\node{}3-\node{\topnode{}2}4-\node{}3-\node{}2-\node{}1$ & $\overline{{\rm min}_{E_7}}$ \\ \hline
	2 & $\orbit{min}{E_8}$ & $\node{}2-\node{}4-\node{\topnode{} 3}6-\node{}5-\node{}4-\node{}3-\node{\squ 1}2\ $ & &\\ \hline
	3 & $\orbit{n.n.min}{E_7}$ & $\node{}2-\node{}4-\node{\topnode{}3}6-\node{}5-\node{}4-\node{\squ 2}3$ & ${\color{white}\node{}1-\node{\topnode{}1}2-\node{}2-\node{\squ 1}2-}\node{\squ 2}1 $ & $\orbit{min}{A_1}$ \\ \hline
	4 & $\orbit{n.min}{E_7}$ & $\node{}2-\node{}4-\node{\topnode{}3}6-\node{}5-\node{\squ 1}4-\node{}2$ & $\node{}1-\node{\topnode{}1}2-\node{}2-\node{\squ 1}2-\node{}1 $ & $\orbit{min}{D_6}$ \\ \hline
	5 & $\orbit{min}{E_7}$ & $\node{\squ 1}2-\node{}3-\node{\topnode{}2}4-\node{}3-\node{}2-\node{}1$ &  &  \\ \hline
	6 & $\orbit{n.min}{E_6}$ & $\node{\squ 1}2-\node{}3-\node{\topnode{}2}4-\node{}3-\node{\squ 1}2$ & $\node{\squ 1}1-\node{}1-\node{}1-\node{}1-\node{\squ 1}1 $ & $\orbit{min}{A_5}$ \\ \hline
	7 & $\orbit{min}{E_6}$ & $\node{}1-\node{}2-\node{\topnode{\squ 1}2}3-\node{}2-\node{}1$ & &  \\ \hline
	8 & $\orbit{n.min}{F_4}$ & $\node{}2-\node{}4\Rightarrow\node{}3-\node{\squ 1}2$ & $\node{}1\Rightarrow\node{}1-\node{\squ 1}1$& $\orbit{min}{C_3}$  \\ \hline
	9 & $\orbit{min}{F_4}$ & $\node{\squ 1}2-\node{}3\Rightarrow\node{}2-\node{}1$ & & \\ \hline
	\end{tabular}
	\caption{Kraft-Procesi transitions between closures of nilpotent orbits of exceptional Lie algebras. $\orbit{min}{G}$ denotes the closure or the minimal nilpotent orbit of $Lie(G)$. Similarly, we use $\orbit{n.min}{G}$ for the next to minimal nilpotent orbit and $\orbit{n.n.min}{G}$ for the next orbit in dimension after the next to minimal. The subtraction quiver $\dif_i\define \q_i-\q_{i+1}$ is computed and its Coulomb branch $\Cb{\dif_i}$ corresponds to the transverse slice between $\q_i$ and a point in the interior of $\q_{i+1}$.}
	\label{tab:excep}
\end{table}

\paragraph{Quiver Subtraction in Exceptional Algebras} Given the results of \cite[fig.~1]{Hanany:2017ooe}, we have a set of quivers whose Coulomb branches are closures of nilpotent orbits of exceptional Lie algebras with \emph{height}\footnote{The \emph{height} of a nilpotent orbit $\hto$ is defined as in \cite[sec. 2]{Panyushev:1999on}. The closure of a nilpotent orbit of height $\hto\leq 2$ can always be realized as the Coulomb branch of a $3d\ \mathcal{N}=4$ quiver based on the \emph{weighted Dynkin diagram} of the orbit (see more details in the following section). We thank Giulia Ferlito for pointing this out.} $\hto = 2$. The quivers of this set corresponding to the groups $F_4$, $E_6$, $E_7$ and $E_8$ satisfy the properties for quiver subtractions (Sec. \ref{sec:def}), and hence their Coulomb branches can be studied with this new procedure. Table \ref{tab:excep} contains the result of subtracting quivers whose Coulomb branches are closures of nilpotent orbits related by a link in their corresponding Hasse diagrams. The Coulomb branches of the subtraction quiver $\dif_i$ correspond indeed with the expected transverse slices \cite{FJLS15}. Therefore, the Kraft-Procesi transition (that in the mathematical literature was introduced in \cite{KP81,KP82} and extended to exceptional algebras by \cite{FJLS15}) has a physical realization also for exceptional algebras (beyond the classical realization in \cite{CH16,CH17}). Notice that the case of $F_4$ constitutes the first example of a Kraft-Procesi transition between non simply laced quivers. These types of quivers and their Coulomb branches were introduced in \cite{Cremonesi:2014xha} during the analysis of moduli spaces of $F_4$, $G_2$, $B_n$ and $C_n$ instantons.

The physical implications are as follows. Let us consider the subtraction $\dif_i=\q_i-\q_{i+1}$. Then, a subset of the Coulomb branch of $\q_i$ is isomorphic to $\Cb{\q_{i+1}}$. If one restricts the vacuum expectation values of the corresponding massless vector multiplets in $\Cb {\q_i}$ to be in the subset ${\Cb{\q_{i+1}}}^\circ\subset \Cb {\q_i}$, the VEVs of the remaining massless vector multiplets sit at a singular point in the directions perpendicular to ${\Cb{\q_{i+1}}}^\circ$. This singularity is defined by $\Cb{\dif_i=\q_i-\q_{i+1}}$. The fact that there are VEVs sitting at the singular point of $\Cb{\dif_i}$ implies that new massless states appear (including massless hypermultiplets), and with them the possibility of transitioning to a mixed branch in the moduli space, which is equivalent to the transition that takes place at the origin of the Coulomb branch in the theory defined by the quiver $\dif_i$. In this way, the computation of transverse slices allows for a global view on mixed branches. For each case there is a singular transverse slice, we get a phase with mixed Higgs and Coulomb branches.

\paragraph{A New Prediction in $E_7$}
It is worth noting that the quiver subtractions can be utilized to predict a new result, namely the transverse slice between closures of nilpotent orbits that are not linked together in their Hasse diagram. This is the case for $\orbit{n.n.min}{E_7}$, with complex dimension $54$, and $\orbit{min}{E_7}$, with complex dimension $34$. Conjecture \ref{con} predicts that the slice $S\subseteq \orbit{n.n.min}{E_7}$ transverse to ${\rm min}_{E_7}$ is isomorphic to the Coulomb branch of the quiver subtraction $\q_3-\q_5$ (where the labels from Table \ref{tab:excep} are used). Let us compute the subtraction:
\ba 
	\dif 	&= \q_3-\q_5\\
		&=\lb  \node{}2-\node{}4-\node{\topnode{}3}6-\node{}5-\node{}4-\node{\squ 2}3\ \ \rb - \lb \node{\squ 1}2-\node{}3-\node{\topnode{}2}4-\node{}3-\node{}2-\node{}1\rb\\
		&=\node{}0- \node{} 1-\node{\topnode{}1}2-\node{}2-\node{}2-\node{\squ 2}2\\
		&= \node{} 1-\node{\topnode{}1}2-\node{}2-\node{}2-\node{\squ 2}2\\
\ea
The Coulomb branch of $\dif$ can be seen from the work in \cite{Hanany:2016gbz} to be the closure of the next to minimal nilpotent orbit\footnote{This is the same as the Higgs branch of the $3d \ \mathcal N=4$ orthosymplectic quiver $\sqnode{}{O(12)}-\node{}{USp(2)}-\node{}{O(1)}$.} of $\mathfrak{so} (12)$, with complex dimension 20:
\be 
	\Cb \dif = \lb \node{} 1-\node{\topnode{}1}2-\node{}2-\node{}2-\node{\squ 2}2\ \ \rb = \orbit{n.min}{D_6}
\ee
Hence, according to Conjecture \ref{con}, the slice $S\subseteq \orbit{n.n.min}{E_7}$ transverse to ${\rm min}_{E_7}$ is\footnote{The authors would like to thank Paul Levy for confirming that this is indeed the correct transverse slice.} $\orbit{n.min}{D_6}$.

\subsection{Classical Algebras}

The Kraft-Procesi transitions described in \cite{CH17} for Lie algebras of type $B$, $C$ and $D$ where based on \emph{orthosymplectic quivers} and their embedding in Type IIB string theory, introduced in \cite{FH00} and further refined in \cite{GW09,BTX10,CDT13}. However, due to the Barbasch-Vogan \cite{BV85} dual map\footnote{See a particularly clear exposition of this duality map in \cite{A02}.} Higgs or Coulomb branches of this type of quivers can only be closures of nilpotent orbits that are \emph{special}\footnote{This is a result of \cite{CDT13}. See also the discussion in \cite{CH17}.}. Due to this, the Kraft-Procesi transitions in \cite{CH17} are restricted to the set of special nilpotent orbits. In particular, it has not been found an \emph{orthosymplectic quiver} whose Coulomb branch is $b_n$, the closure of the minimal nilpotent orbit of $\mathfrak{so}(2n+1,\mathbb{C})$, or $c_n$, the closure of the minimal nilpotent orbit of $\mathfrak{sp}(n,\mathbb{C})$. In this section we show how the subtraction of quivers, together with Conjecture \ref{con}, can be used to overcome this obstacle, at least for nilpotent orbits of height lower or equal than 2, and provide a physical realization of Kraft-Procesi transitions of type $b_n$ and $c_n$.

\paragraph{Transitions to $b_4$}

Utilizing $3d \ \mathcal N=4$ quivers based on Dynkin diagrams \cite{Cremonesi:2014xha,Hanany:2016gbz}, let $\q _{(2^2,1^5)}$ be the quiver:
\be 
	\q_{(2^2,1^5)} \define \ \node{}{1}-\node{\squ 1}2-\node{}2\Rightarrow\node{}1
\ee

such that its Coulomb branch is the closure of the minimal nilpotent orbit of $\so 9$:
\be 
	\Cb{\q_{(2^2,1^5)}} = \bOr{2^2,1^5}\subset \so 9
\ee
Employing the notation of \cite{KP82} we write $\Cb{\q_{(2^2,1^5)}} =b_4$. There are two nilpotent orbits that sit directly above of $\bOr{2^2,1^5}$ in the Hasse diagram of $\so 9$: $\bOr{2^4,1}$ and $\bOr{3,1^6}$. The $3d\ \mathcal N=4$ quivers $\q_{(2^4,1)}$ and  $\q_{(3,1^6)}$ such that
\begin{align}
	\CC\left(\q_{(2^4,1)}\right) &= \bOr{2^4,1} \subset \so 9 \\
	\CC\left(\q_{(3,1^6)}\right) &= \bOr{3,1^6}\subset \so 9
\end{align} 
are:
\begin{align}
	\q_{(2^4,1)} &\define\ \node{}{1}-\node{}2-\node{}3\Rightarrow\node{\squ 1}2 \\
	\q_{(3,1^6)} &\define\ \node{\squ 2}{2}-\node{}2-\node{}2\Rightarrow\node{}1 
\end{align}
We can perform different subtractions for each quiver, obtaining:
\ba
	\dif_{(2^4,1)} &= \q_{(2^4,1)} - \q_{(2^2,1^5)}\\
		&= \left( \node{}{1}-\node{}2-\node{}3\Rightarrow\node{\squ 1}2\ \ \right)- \left(\node{}{1}-\node{\squ 1}2-\node{}2\Rightarrow\node{}1\right)\\
		&=\ \node{}0-\node{}0-\node{}1\Rightarrow\node{\squ 1}1\\
		&=\ \node{}1\Rightarrow\node{\squ 1}1
\ea
and
\ba
	\dif_{(3,1^6)} &= \q_{(3,1^6)} - \q_{(2^2,1^5)}\\
		&= \left( \node{\squ 2}{2}-\node{}2-\node{}2\Rightarrow\node{}1  \right)- \left(\node{}{1}-\node{\squ 1}2-\node{}2\Rightarrow\node{}1\right)\\
		&=\ \node{\squ 2}1-\node{}0-\node{}0\Rightarrow\node{}0\\
		&=\ \node{\squ 2}1
\ea
The Coulomb branches of the subtraction quivers are:
\begin{align}
	\Cb {\dif _{(2^4,1)}} &= b_2\\
	\Cb {\dif _{(3,1^6)}} &= a_1
\end{align}
where $b_2$ is the closure of the minimal nilpotent orbit of $\so 5$ and $a_1$ is the closure of the minimal nilpotent orbit of $\sl 2$. According to Conjecture \ref{con}, $\Cb {\dif _{(2^4,1)}}$ and  $\Cb {\dif _{(3,1^6)}}$ should be the transverse slices between the corresponding orbits and the minimal orbit of $\so 9$. This is indeed the case, as it was computed by Kraft and Procesi in \cite{KP82}. 

\paragraph{Generalization} This physical realization can be generalized to transverse slices between nilpotent orbits of height $\text{ht}(\mathcal O)\leq 2$ of any Lie algebra. The $3d \ \mathcal N=4$ quiver $\q $ with the Coulomb branch $\Cb {\q}=\bar{\mathcal O}$ being the closure of the nilpotent orbit $\mathcal O$ is always formed from the \emph{weighted Dynkin diagram} \cite{CM93} of the corresponding orbit\footnote{Note that the labels in the \emph{weighted Dynkin diagram} of nilpotent orbits of classical Lie algebras coincide with the \emph{Root Maps} tabulated in \cite[App.~B]{Hanany:2016gbz}. In \cite{Hanany:2017ooe} the labels of the \emph{weighted Dinkin diagram} are called the \emph{Characteristics} of the orbit.}. The nodes in the weighted Dynkin diagram become gauge nodes in the quiver (with the same connections as the Dynkin diagram), the labels in the weighted Dynkin diagram become the ranks of flavor nodes attached to each different gauge node, and the ranks of the gauge nodes are determined by imposing that all of them should be balanced (i.e. number of flavors of the node is equal to twice its rank).

Let us illustrate the result with one further example from a $C$ type algebra and one from a $D$ type. 

\paragraph{C type transition}

Let the quivers $\q_{(2^2,1^4)}$ and $\q_{(2,1^6)}$ be:
\begin{align}
	\q_{(2^2,1^4)} &\define\ \node{}1-\node{\squ 1}2-\node{}2\Leftarrow\node{}2  \\
	\q_{(2,1^6)} &\define\ \node{\squ 1}1-\node{}1-\node{}1\Leftarrow\node{}1
\end{align}
Their Coulomb branches are:
\begin{align}
	\Cb{\q_{(2^2,1^4)}}&=\bOr{2^2,1^4}\subset \sp 4\\
	\Cb{\q_{(2,1^6)}}&=\bOr{2,1^6}\subset \sp 4
\end{align}
The subtraction can be computed:
\ba
	\dif_{(2^2,1^4)} &= \q_{(2^2,1^4)} - \q_{(2,1^6)}\\
		&= \left(\node{}1-\node{\squ 1}2-\node{}2\Leftarrow\node{}2  \right)- \left(\node{\squ 1}1-\node{}1-\node{}1\Leftarrow\node{}1 \right)\\
		&=\ \node{}0-\node{\squ 1}1-\node{}1\Leftarrow\node{}1\\
		&=\ \node{\squ 1}1-\node{}1\Leftarrow\node{}1\\
\ea
Its Coulomb branch is
\be 
	\Cb{\dif_{(2^2,1^4)}}=\Cb{\node{\squ 1}1-\node{}1\Leftarrow\node{}1}= c_3
\ee
where $c_3=\bOr{2,1^4}\subset \sp 3$ is the closure of the minimal nilpotent orbit of $\sp 3$. $c_3$ is indeed the transverse slice between the two orbits \cite{KP82}.

\paragraph{D type transition}

Let the quivers $\q_{(2^4,1^2)}$ and $\q_{(2^2,1^6)}$ be:
\begin{align}
	\q_{(2^4,1^2)} &\define\ \node{}1-\node{}2-\node{\topnode{\squ 1}2}3-\node{\squ 1}2  \\
	\q_{(2^2,1^6)} &\define\ \node{}1-\node{\squ 1}2-\node{\topnode{}1}2-\node{}1
\end{align}
Their Coulomb branches are:
\begin{align}
	\Cb{\q_{(2^4,1^2)}}&=\bOr{2^4,1^2}\subset \so {10}\\
	\Cb{\q_{(2^2,1^6)}}&=\bOr{2^2,1^6}\subset \so {10}
\end{align}
The subtraction can be computed:
\ba
	\dif_{(2^4,1^2)} &= \q_{(2^4,1^2)} - \q_{(2^2,1^6)}\\
		&= \left(  \node{}1-\node{}2-\node{\topnode{\squ 1}2}3-\node{\squ 1}2 \ \right)- \left( \node{}1-\node{\squ 1}2-\node{\topnode{}1}2-\node{}1\right)\\
		&=\ \node{}0-\node{}0-\node{\topnode{\squ 1}1}1-\node{\squ 1}1\\
		&=\ \node{\topnode{\squ 1}1}1-\node{\squ 1}1\\
		&=\ \node{\squ 1}1-\node{}1-\node{\squ 1}1  
\ea
Its Coulomb branch is
\be 
	\Cb{\dif_{(2^4,1^2)}}=\Cb{\node{\squ 1}1-\node{}1-\node{\squ 1}1 \ \ }= d_3
\ee
where $d_3=\bOr{2^2,1^2}\subset \so 6$ is the closure of the minimal nilpotent orbit of $\so 6$. $d_3$ is indeed the transverse slice between the two orbits \cite{KP82}. Therefore, the physical realization of Kraft-Procesi transitions is extended in this way to quivers that do not necessarily have an embedding in Type IIB string theory.

\section{Integrating Out a Massive Quark in $5d$ at infinite coupling}\label{sec:5d}

This section analyses the physics of integrating out massive quarks in $5d\ \mathcal N=1$ (eight supercharges) SQCD theories, at the limit $1/g^2\rightarrow 0$, where $g$ is the gauge coupling.  Let $\h_9$ be the quiver of one of these $5d$ theories with gauge group $G=SU(3)$, nine quarks $N_f=9$, and Chern-Simons level $k=1/2$:  
\be 
	\h_9 \define \  \node{\flav 9}{SU(3),~1/2}
\ee 
Its Higgs branch at infinite coupling $1/g^2\rightarrow 0$ is isomorphic to the Coulomb branch of a $3d \ \mathcal N=4$ quiver \cite{FHMZ17}. Let us recycle the labels $\q_i$ from the previous section and define the $3d$ quiver $\q_9$  as:
\be 
	\q_9 \define \ \node{}{1}- \node{}{2}- \node{}{3}- \node{}{4}- \node{}{5}- \node{}{6}- \node{}{7}- \node{\upnode{4}}{8}- \node{}{5}- \node{}{2}
\ee
The relation can formally be written as:

\be 
	\CH_\infty(\h_9) = \CC^{3d} (\q_9)
\ee
In the following, the label $\h_i$ is reserved for quivers of $5d$ theories and $\q_i$ and $\dif_i$ for $3d$ quivers. If a massive quark is integrated out from $\h_9$ this will result in a new effective theory with one less flavor $N_f=8$, the same gauge group $G=SU(3)$, and a Chern-Simons level increased by $1/2$, $k=1$. Let the resulting theory's quiver be $\h_8$:
\be 
	\h_8 \define \  \node{\flav 8}{SU(3),~1}
\ee 

One can ask the question \emph{what is the effect of integrating out this massive quark on the Higgs branch at infinite coupling?} In order to answer this let us write the quiver $\q_8$:
\be 
	\q_8	\define \node{}{1}- \node{}{2}- \node{}{3}- \node{}{4}- \node{}{5}- \node{\upnode{3}}{6}- \node{}{4}- \node{}{2}- \node{}{1}
\ee
such that \cite{FHMZ17}:
\be 
	\CH_\infty(\h_8) = \CC^{3d} (\q_8)
\ee
Then, we have
\be 
	\CH_\infty(\h_8)\subset \CH_\infty(\h_9)
\ee
and the effect of a massless quark becoming massive, while the gauge coupling is infinite, can be identified utilizing the $3d$ description $\CC^{3d} (\q_8)\subset \CC^{3d} (\q_9)$.  
Let us denote the quiver resulting from the quiver subtraction by:
\be
\mathsf {\dif}_9 = \q_9-\q_8
\ee
The first thing to notice when one tries to perform the difference $\dif_9$ is that $\q_9$ and $\q_8$ do not have the same number of gauge nodes. However, this can be solved by adding gauge nodes with label zero to $\q_8$. Note that the rightmost node of $\q_8$ with label $1$ still constitutes a problem, since adding an extra node to the right of $\q_9$ with label zero and then subtracting would result in a negative label $-1$. Instead, one can use the decoupling of a $U(1)$ from the gauge group of a $3d\ \mathcal N=4$ quiver with no flavor nodes to rewrite:
\be 
	\q_8	=\  \node{}{1}- \node{}{2}- \node{}{3}- \node{}{4}- \node{}{5}- \node{\upnode 3}{6}- \node{}{4}- \node{\flav 1}{2}
\ee
Let us also decouple a $U(1)$ from the leftmost node in $\q_9$, and add one extra zero node to the left of $\q_8$. Then, the two $3d$ quivers satisfy the conditions for \emph{quiver subtraction} (Sec. \ref{sec:def}); they are:
\ba 
	\q_9 &= \  \node{\flav 1}{2}- \node{}{3}- \node{}{4}- \node{}{5}- \node{}{6}- \node{}{7}- \node{\upnode{4}}{8}- \node{}{5}- \node{}{2}\\
	\q_8	&=\  \node{}{0}-  \node{}{1}- \node{}{2}- \node{}{3}- \node{}{4}- \node{}{5}- \node{\upnode 3}{6}- \node{}{4}- \node{\flav 1}{2}
\ea
Now the difference can be taken:
\ba 
	\dif_9	&= \q_9-\q_8\\
		&= \lb \node{\flav 1}{2}- \node{}{3}- \node{}{4}- \node{}{5}- \node{}{6}- \node{}{7}- \node{\upnode{4}}{8}- \node{}{5}- \node{}{2} \rb -\\
		&- \lb \node{}{0}-  \node{}{1}- \node{}{2}- \node{}{3}- \node{}{4}- \node{}{5}- \node{\upnode 3}{6}- \node{}{4}- \node{\flav 1}{2} \rb \\
		&=\  \node{\flav 1}{2}- \node{}{2}- \node{}{2}- \node{}{2}- \node{}{2}- \node{}{2}- \node{\upnode{1}}{2}- \node{}{1}- \node{}{0} \\
		&=\  \node{\flav 1}{2}- \node{}{2}- \node{}{2}- \node{}{2}- \node{}{2}- \node{}{2}- \node{\upnode{1}}{2}- \node{}{1} \\
\ea 

The Coulomb branch of $\dif_9$ is known to be freely generated by twisted hypermultiplets:
\be
	\CC^{3d}(\dif_9) = \CC^{3d}\lb \node{\flav 1}{2}- \node{}{2}- \node{}{2}- \node{}{2}- \node{}{2}- \node{}{2}- \node{\upnode{1}}{2}- \node{}{1} \rb = \mathbb{C}^{32}
\ee

One can use this result to describe the \emph{``slice"} $S\subseteq \CH_\infty(\h_9)$ \emph{``transverse"} to a point in the interior of $\CH_\infty(\h_8)$ as:
\be 
	S=\co ^{32} = \qu ^{16}
\ee
Note that in this case this is not a \emph{transverse slice} between nilpotent orbits, but the \emph{quiver subtraction} can extend the notion to quivers whose Coulomb branches are not closures of nilpotent orbits. Another difference with the previous cases studied here is that the moduli space $S=\CC^{3d}({\dif_9})=\qu ^{16}$ is not singular. In this case, if the VEVs of some vector multiplets of $\q_9$ are tuned to restrict $\CC^{3d}({\q_9})$ to its subset $\c3d{\q_8}^\circ$, the VEVs that are tuned are not in a singular point with respect to the directions in $\c3d{\q_9}$ perpendicular to the subset $\c3d{\q_8}^\circ$. This means that there are no new massless states arising as a consequence of this tuning, and there are no new branches opening up. From the point of view of the $5d$ theory, this means that the process of integrating out a massive quark, while the gauge coupling remains infinite, does not involve the creation of new massless states.

One can proceed integrating out consecutive massive quarks from the theory until the last quiver $\h_5$ is found:
\be 
	\h_5 \define \node{\flav 5}{SU(3),~5/2}
\ee

The effects can be computed by employing quiver differences in a way analogous to $\dif_9$, and are presented in Table \ref{tab:5d9}. Note that the transverse spaces $\CC^{3d}(\dif_i)$ are always freely generated by twisted hypermultiplets, and hence have the form $\qu ^n$. This implies that the physics of integrating out a massive quark in $5d$ SQCD with eight supercharges does not include the appearance of new massless states.

\begin{table}[t]
	\centering
	\noindent\begin{tabularx}{\linewidth}{|c|c|X|X|c|}
		\hline
		$i$ & $\h_i$ & \multicolumn{1}{c|}{$\q_i$} & \multicolumn{1}{c|}{$\dif_i$} & $\CC^{3d}( {\dif_i})$  \\ \hline
		$9$ & $\node{\flav 9}{SU(3),~1/2}$ & $ \node{\squ 1}{2}- \node{}{3}- \node{}{4}- \node{}{5}- \node{}{6}- \node{}{7}- \node{\topnode{}{4}}{8}- \node{}{5}- \node{}{2}$ & $\node{\squ 1}{2}- \node{}{2}- \node{}{2}- \node{}{2}- \node{}{2}- \node{}{2}- \node{\topnode{}{1}}{2}- \node{}{1}{\color{white}-\node{}{0}} $ & $\qu^{16}$ \\ \hline
		$8$ & $\node{\flav 8}{SU(3),~1}$ & ${\color{white}\node{}{0}-}\node{}{1}- \node{}{2}- \node{}{3}- \node{}{4}- \node{}{5}- \node{\topnode{}{3}}{6}- \node{}{4}- \node{\squl 1}{2}$ & ${\color{white}\node{}{0}-}\node{}{1}- \node{}{1}- \node{}{1}- \node{}{1}- \node{}{1}- \node{}{1}- \node{}{1}- \node{\squl 1}{1}$ & $\qu^{8}$\\ \hline	
		$7$ & $\node{\flav 7}{SU(3),~3/2}$ & ${\color{white}\node{}0-\node{}0-}\node{}{1} - \node{}{2} - \node{}{3} - \node{}{4}-\node{\topnode{\squ 1}{3}}{5}- \node{}{3} - \node{}{1}$ & ${\color{white}\node{}0-\node{}0-}\node{}1- \node{}1- \node{}1- \node{}1- \node{\topnode{\squ {1}}{1}}1{\color{white}-\node{}{0}-\node{}{0}}$ & $\qu^{6}$\\ \hline	
		$6$ & $\node{\flav 6}{SU(3),~2}$ & ${\color{white}\node{}0-\node{}0-\node{}0-}\node{}{1} - \node{} 2 -\node{}3 -\node{\topnode{} 2} 4-\node{\squ 1} 3 -\node{} 1$ & ${\color{white}\node{}0-\node{}0-\node{}0-}\node{}1-\node{}1-\node{}1-\node{}1-\node{\squ 1}1{\color{white}-\node{}{0}}$ & $\qu^{5}$\\ \hline	
		$5$ & $\node{\flav 5}{SU(3),~5/2}$ & ${\color{white}\node{}0-\node{}0-\node{}0-\node{}0-}\node{}{1} - \node{}{2}-\node{\topnode{\squ 1}2}3-\node{}2 - \node{\flav 1} 1 $ & & \\ \hline	
	\end{tabularx}
	\caption{Quiver differences between Higgs branches of $5d$ theories at infinite coupling. $\h_i$ are $5d$ quivers. $\q_i$ are $3d$ quivers such that $\CH_{\infty}(H_i)=\CC^{3d}(\q_i)$ \cite{FHMZ17}. On the right, the quiver subtractions have been computed $\dif_i=\q_{i}-\q_{i-1}$.}
	\label{tab:5d9}
\end{table}
The same analysis can be done with different starting points:
\begin{align} 
		\iq_8 &\define \ \node{\flav 8}{SU(3),0}\\
		\jq_6 &\define \ \node{\flav 6}{SU(3),0}
\end{align}
The results of integrating out massive quarks one by one are depicted on tables \ref{tab:5d8} and \ref{tab:5d6} respectively.

\begin{table}[t]
	\centering
	\noindent\begin{tabularx}{\linewidth}{|c|c|X|X|c|}
		\hline
		$i$ & $\iq_i$ & \multicolumn{1}{c|}{$\q_i$} & \multicolumn{1}{c|}{$\dif_i$} & $\CC^{3d}( {\dif_i})$  \\ \hline
		$8$ & $\node{\flav 8}{SU(3),~0}$ & $\node{\squ 1}{2}- \node{}{3}- \node{}{4}- \node{\topnode{} 2}{5}- \node{}{4}- \node{}{3}- \node{}{2}- \node{}{1}$ &$ \node{\squ 1}{1}- \node{}{1}- \node{}{1}- \node{}{1}- \node{}{1}- \node{}{1}- \node{}{1}- \node{}{1}$  &$\qu^{8}$ \\ \hline	
		$7$ & $\node{\flav 7}{SU(3),~1/2}$ & $\node{}{1} - \node{}{2} - \node{}{3} - \node{\topnode{\squ 1} 2}{4}- \node{}{3} - \node{}{2}-\node{}1$ & ${\color{white}\node{}{}-\node{}{}-\node{}{}-} \node{\topnode{\squ{1}}{1}}1 - \node{}1- \node{}1- \node{}1$& $\qu^{5}$\\ \hline	
		$6$ & $\node{\flav 6}{SU(3),~1}$ & $\node{}{1} - \node{} 2 -\node{\squ1}3 -\node{\topnode{}1} 3-\node{} 2 -\node{} 1$ &  ${\color{white}\node{}{}-\node{}{}-}\node{\squ 1}1-\node{}1-\node{}1-\node{}1$& $\qu^{4}$\\ \hline	
		$5$ & $\node{\flav 5}{SU(3),~3/2}$ & $\node{}{1}- \node{\squ 1}{2}-\node{}2-\node{\topnode{\squ 1} 1}2-\node{}1$ & & \\ \hline	
	\end{tabularx}
	\caption{Quiver differences between Higgs branches of $5d$ theories at infinite coupling. $\iq_i$ are $5d$ quivers. $\q_i$ are $3d$ quivers such that $\CH_{\infty}(\iq_i)=\CC^{3d}(\q_i)$ \cite{FHMZ17}. On the right, the quiver subtractions have been computed $\dif_i=\q_{i}-\q_{i-1}$.}
	\label{tab:5d8}
\end{table}

\begin{table}[t]
	\centering
	\noindent\begin{tabularx}{\linewidth}{|c|c|X|X|c|}
		\hline
		$i$ & $\jq_i$ & \multicolumn{1}{c|}{$\q_i$} & \multicolumn{1}{c|}{$\dif_i$} & $\CC^{3d}( {\dif_i})$  \\ \hline
		$6$ & $\node{\flav 6}{SU(3),~0}$ & $ \node{}{1}-\node{}{2}-\node{\overset{{\llap{$\overset{{}}{\overset{{1}}{\scriptstyle\square}} -$}\displaystyle \overset{{2}}{\circ}{\rlap{$-\overset{{}}{\overset{{1}}{\circ}}$}}}}{\scriptstyle\vert}}{3} -\node{}{2}-\node{}{1} $ & $ {\color{white}\node{}{0}-\node{}{0}-}\node{\overset{{\llap{$\overset{{}}{\overset{{1}}{\scriptstyle\square}} -$}\displaystyle \overset{{1}}{\circ}}}{\scriptstyle\vert}}{1} -\node{}{1}-\node{}{1} $ & $\qu^{4}$\\ \hline	
		$5$ & $\node{\flav 5}{SU(3),~1/2}$ & $\node{}1-\node{\flav 1}2-\node{\overset{{\displaystyle \overset{{1}}{\circ}{\rlap{$-{\overset{{1}}{\circ}}-{\overset{{1}}{\scriptstyle \square}}$}}}}{\scriptstyle\vert}}{2}-\node{}1$ & & \\ \hline	
	\end{tabularx}
	\caption{Quiver differences between Higgs branches of $5d$ theories at infinite coupling. $\jq_i$ are $5d$ quivers. $\q_i$ are $3d$ quivers such that $\CH_{\infty}(\jq_i)=\CC^{3d}(\q_i)$ \cite{FHMZ17}. On the right, the quiver subtractions have been computed $\dif_i=\q_{i}-\q_{i-1}$.}
	\label{tab:5d6}
\end{table}

\section{Conclusions and Outlook}\label{sec:con}

We have formally defined the concept of \emph{quiver subtractions}. This idea is a direct consequence of the physical realization of \emph{transverse slices} and \emph{Kraft-Procesi transitions} in terms of D-branes dynamics in Type IIB string theory and moduli spaces of $3d\ \mathcal N=4$ quiver gauge theories \cite{CH16}. However, it has recently been discovered to be a more general construction, that plays an important role on the description of $6d\ \mathcal N=(1,0)$ physics, i.e., the small $E_8$ instanton transition \cite{HM18}. In this note we provide an example in which it can be utilized to analyze the physics at infinite gauge coupling of $5d\ \mathcal N=1$ SQCD theories. In particular we show how to characterize the effect of integrating out a massive quark on the Higgs branch of the theory at infinite coupling. 

We have also shown explicitly how the \emph{quiver subtraction} extends the physical realization of transverse slices and Kraft-Procesi transitions in nilpotent orbits of classical algebras \cite{CH16,CH17} to exceptional Lie algebras and to nilpotent orbits of classical algebras that are non-special. We want to mention once more that the tool presented herein is remarkably simple. We hope that the examples shown are sufficiently illuminating of its computational power. We believe that these two features will make of \emph{quiver subtractions} a very useful tool in the set of resources available to study the physics of $6d$ and $5d$ theories with eight supercharges. One computation that remains to be done, for which \emph{quiver subtractions} are specially suited, is the transverse slice between the Higgs branch of $5d \ \mathcal N=1$ SQCD at finite and at infinite coupling. It is also desirable to produce a similar formal description of quiver subtractions for orthosymplectic quivers, derived from the work in \cite{CH17} and also utilized in \cite{HM18}.

\section*{Acknowledgments}
We would like to thank Giulia Ferlito, Rudolph Kalveks, Paul Levy, Noppadol Mekareeya, Claudio Procesi, Travis Schedler and Gabi Zafrir for helpful conversations during the development of this project. A.H. is thankful for the hospitality of Aspen Center for Physics and the organizers of the Winter Conference \emph{Superconformal Field Theories
in Four or More Dimensions} where the ideas for this work where kindled. The two authors are also thankful for the hospitality of the IFT in Madrid and the organizers of the conference \emph{Physics and Geometry of F-Theory}, were some of this project was further developed. S.C. is supported by an EPSRC DTP studentship EP/M507878/1. A.H. is supported by STFC Consolidated Grant ST/J0003533/1, and EPSRC Programme Grant EP/K034456/1.

\bibliography{main}
\bibliographystyle{JHEP}

\end{document}